\documentstyle[12pt]{article}    
\begin{document}           
\baselineskip=0.33333in
\begin{quote} \raggedleft TAUP 2806-2005
\end{quote}
\vglue 0.5in
\begin{center}{\bf Relativistic Constraints on the Structure \\
of Fundamental Forces}

\end{center} 
\begin{center}E. Comay 
\end{center}
 
\begin{center}
School of Physics and Astronomy \\
Raymond and Beverly Sackler Faculty of Exact Sciences \\
Tel Aviv University \\
Tel Aviv 69978 \\
Israel
\end{center}

Email: elic@tauphy.tau.ac.il
\vglue 0.5in
\vglue 0.5in
\noindent
\vglue 0.2in
Abstract:  

It is proved that Special Relativity imposes constraints on the structure
of fundamental forces. The orthogonality of the 4-force exerted on an
elementary particle and its 4-velocity is discussed. The significance
of the energy-momentum tensor associated with the field is analyzed. 
Relying on these issues, it is
proved that the Lorentz force is consistent with all constraints whereas
a force derived from a scalar potential does not satisfy all
requirements. This analysis explains a general discussion of
Goldstein, Poole and Safko.

\newpage
\noindent
{\bf 1. Introduction}
\vglue 0.33333in

It is well known that the theory of Special Relativity (SR) changed
dramatically old
concepts concerning the structure of the physical world. The following
lines mention several examples of this kind. Thus, probably the most
well known relativistic formula $E=mc^2$ unifies mass and energy
conservation laws. (In this formula $m$ denotes the mass measured in
the laboratory
frame. In all other cases, $m$ is a scalar denoting the self mass of a
particle.) Another notion is the absolute property ascribed to
length and time intervals. This idea has been forsaken and a single absolute
quantity $(ds)^2 = (dt)^2 - (d{\bf r})^2$ 
is defined. Velocity takes a new form. In
SR, velocity of a massive particle is always smaller than
the speed of light $v < c$. In the case of massless particles, like the
photon, the theory claims that it travels in the speed of light $c$ in
all inertial frames and that $c$ is independent of the velocity of the source. 
The introduction of SR into
quantum mechanics enabled Dirac to construct his celebrated equation. Solutions
of this equation yield very good values for the atomic hydrogen energy
levels, for the spin of the electron and for its g-factor. 
Another result of this equation is
the prediction of the positron - an antiparticle of the electron.

SR shows that one does not need to
postulate the existence of ether in order to explain wave properties of
electromagnetic radiation.  Thus, the following 
very well known textbooks on electrodynamics
[1-3] do not have the words "ether" or "aether" in the subject index. 
The very detailed book of Jackson
(see [4], pp. 503-515) discusses ether and other ideas which are
inconsistent with SR and describes relevant experiments aiming to test these ideas.
All results are consistent with SR. 

Another difficulty of the ether theory emerges from quantum mechanical
effects. Experiments proves and this theory explains the particle/wave
duality of other elementary particles, like the electron as well as
that of nonelementary objects like the proton and the neutron. Thus, if
one postulates that for every wave there should be a material medium 
(analogous to the case of acoustic waves) then 
a need for other kinds of ether (or new properties of
the same ether) becomes a necessity. This argument explains the above mentioned
status of ether in the current scientific literature. In spite
of that, the notion of ether still can be found in modern
scientific and nonscientific texts. Thus,
a search of the Internet by means of Google shows about 1000 cases for
the strings "ether waves", "aether waves". This figure should be compared
with the 495,000 cases found for "electromagnetic waves".

SR changed the notions of probability (or charge) density and of energy
density. Thus, in Newtonian mechanics volume is invariant and so is time.
It follows that, at any instant, 
the amount of physical objects (like charge or mass) 
enclosed inside a given volume, takes the same value
in all inertial frames. For this reason, density is an invariant in
Newtonian mechanics. This property does not hold in SR because
volumes undergo a Lorentz contraction and simultaneity of events
is not conserved. Thus, it is proved that
probability density (and charge density)
become 0-component of the current 4-vector (see [1], pp. 69-71).
This 4-vector is parallel to the local 4-velocity.

In the case of energy density, one finds that it 
is represented by the $T^{00}$ component of
the energy-momentum tensor $T^{\mu \nu}$. Thus, energy density, momentum
density, energy current and the $3\times 3$ tensor (like the Maxwell stress
tensor of electrodynamics) are entries of one tensor $T^{\mu \nu }$
(see [1], pp. 77-83). A Lorentz transformation of this tensor 
indicates the close relations between these notions.

SR explains a tremendous amount of data. Perhaps the most
striking case is found in the colliding electron/positron beams of the
LEP accelerator at CERN. Here electrons and positrons acquire a kinetic
energy larger than 100GeV. It means that $E_K > 200,000\,m_ec^2$, where
$m_e$ denotes the rest mass of the electron (and of the positron). Now, in
spite of this gigantic kinetic energy, the velocity of the particles
does not exceed the speed of light $c$. Moreover, 
for these beams, an electron-positron
interaction yields plenty of massive particles, demonstrating the
interrelation between mass and energy, which is
expressed by the formula $E=mc^2$. All kinds of processes
like this abide by the law where an elementary particle and its
antiparticle are produced in pairs and the overall energy and momentum
(calculated by the laws of SR) are conserved.

The present work is devoted to the analysis of
relativistic constraints imposed on the
structure of fundamental forces. A brief discussion of the 
special case of the Lorentz force and
of 2 approaches to the problem can be found in a recent edition of a well
known textbook (see [5], pp. 297-300). This discussion indicates that the 
status of the problem is still indecisive.

In the notation used here Greek indices run from 0 to 3. The Lorentz metric $g_{\mu \nu}$
is diagonal and its entries are (1,-1,-1,-1). $\tau $ denotes the proper time. In the system of units used here $\hbar = c=1$ and the 
dimensions of every quantities is a power
of the length $L$. Thus, mass, energy and momentum take the dimension 
$[L^{-1}]$ whereas length and time have the dimension $[L]$.

\vglue 0.666667in
\noindent
{\bf 2. The Notion of Force}
\vglue 0.33333in

People are aware of several kinds of forces encountered in everyday life,
like the force exerted by an extended (or contracted) spring or
other objects obeying Hooke's law; the force
exerted by the pressure of gas on the surface of a piston; several kinds
of friction forces; forces associated with biological activity of muscles
etc. These forces are treated in textbooks where appropriate
formulas are used. However, these kinds of forces do not have a fundamental
nature and their description applies phenomenological formulas. The
phenomenological nature of these forces
is explained in the following lines.

In theoretical physics a genuine elementary particle is pointlike. This
property holds in classical physics (see [1], pp. 43-44).  Here the authors use
several arguments that rely on SR. One of their arguments assumes that
an elementary particle
having a nonzero volume exists. As an elementary particle, no relative motion between
its parts can take place. Now assume that at a certain instant
$t$ a force is exerted
on one of its sides. By the elementary nature of the particle, all its parts
must
move at the same speed. Hence, at the volume of this particle, interaction propagates
at an infinite speed. This result contradicts SR. 

Similarly, the pointlike nature of elementary particles is obtained
in quantum mechanics and in quantum field theory,
where a genuine elementary particle is described
by a wave function (field function) $\psi (x^\mu )$.  
This function depends on a {\em single}
set of space-time coordinates $x^\mu $. Hence, $\psi (x^\mu )$  describes
a pointlike particle.

The pointlike nature of genuine elementary particles, like electrons, muons
and quarks is consistent with experimental data which indicate that their
size is smaller than $10^{-16}\;cm$. This limit depends on the energy used in the
experiments. Today it is believed that genuinely elementary particles like the
electron, the muon and the quarks are pointlike.
(The discussion carried out below uses both pointlike expressions like $ev^\mu$
and density expressions like the 4-current $j^\mu $ (see [1], pp. 69-71). Using
the Dirac $\delta $-function and performing an integration, one
derives the first kind of expression from the second one.)

Evidently, pointlike particles cannot collide. Hence, a fundamental
force cannot stem from a contact interaction between 2 particles.
It follows that a mediating field $F$ is required for explaining 
acceleration of a particle as well as energy 
and momentum exchanged between interacting particles. 
Thus, the rest of this work is
devoted to the analysis of the structure of this field $F$ and to its interrelations
with the force exerted on particles. Here the tensorial nature of $F$ is
still undefined. A discussion of this aspect of the problem is 
presented in the rest of this work. (Hereafter, the word 'collision'
refers to cases where, at the interaction instant, the distance between
the particles is very short.)

The notion of force holds in classical physics where measurements of
position, velocity and acceleration are assumed to yield results having
adequate accuracy. Let us examine the motion of a particle $P$ from
point $A$ to point $B$ along a curve $C$ (see fig. 1). The particle $P$
accelerates and, in the laboratory frame, its energy at $B$ is higher
that that of $A$. Let $\Delta E$ denote the amount of energy acquired
by $P$. Due to the laws of SR, energy cannot travel faster than light.
Hence, when the particle $P$ was at $A$, the energy $\Delta E$ was
inside the spherical shell $S$. But $S$ contains nothing except
the pointlike particle $P$ and the field of force $F$. 
This argument proves that
energy density must be associated with the field of force. The physical
expression for energy density and related quantities
is the energy-momentum tensor $T^{\mu \nu}$
(see [1], pp. 77-80). Here the entry $T^{00}$ represents energy density.
This tensor is a symmetric second rank tensor.

A simple dimensional analysis yields the relations between the field
of force and its associated energy-momentum tensor. Using Newton's law
${\bf f }= m {\bf a}$, one finds that, in the system of units used here,
the dimension of force is $L^{-2}$. Now, energy has the dimension
$L^{-1}$. Therefore, energy density has the dimension $L^{-4}$. These
values restrict the relations between the field
$F$ and $T^{\mu \nu }$.

Another restriction imposed on the system is that, in the vacuum, the
energy-momentum tensor must be divergenceless, $T^{\mu \nu}_{\;\;,\nu} = 0$.
This restriction relies on the fact that the vacuum can be neither
a source nor a sink of energy and momentum.

Before proceeding further, let us examine the case of the electromagnetic force.
This case provides an important illustration of the problem and is used
in the analysis carried out later.

\vglue 0.666667in
\noindent
{\bf 3. The Electromagnetic Force}
\vglue 0.33333in

The electromagnetic force (called the Lorentz force) can be derived
from the action principle. Here we have a charged particle and an
electromagnetic field. The particle's part of the Lagrangian is
(see [1], pp. 45-49)
\begin{equation}
L = -m(1-v^2)^{1/2} - e(\Phi - \bf {A\cdot v}),  
\label{eq:LAGEM}
\end{equation}
where $m$ and $e$ denote the particle's self mass and charge, respectively.
Applying the Euler-Lagrange equations
\begin{equation}
\frac {d}{dt} \frac {\partial L}{\partial \bf {v}} = 
\nabla L
\label{eq:ELEQ}
\end{equation}
to $(\!\!~\ref{eq:LAGEM})$, one isolates the time derivative of the mechanical
momentum and obtains the Lorentz force
\begin{equation}
{\bf f} = \frac {d \bf {p}}{dt} = e(\bf {E} + \bf {v\times B})
\label{eq:LORFORCE}
\end{equation}
This expression can be written in a covariant form
\begin{equation}
f^\mu = ma^\mu = eF^{\mu \nu}v_\nu,
\label{eq:LORF}
\end{equation}
where $v^\mu $ and $a^\mu = dv^\mu /d\tau $ are the particle's 4-velocity and 4-acceleration, respectively. Now, since $v^\mu $ is dimensionless, one
finds that the electromagnetic fields
and the Lorentz force $f^\mu $ of $(\!\!~\ref{eq:LORF})$ have the same
dimensions $[L^{-2}]$.

The energy-momentum tensor of the electromagnetic fields is
(see [1], p. 81 or [4], p. 605))
\begin{equation}
T_{F}^{\mu \nu } =
\frac {1}{4\pi }(F^{\mu \alpha }F^{\beta \nu }g_{\alpha \beta }
+\frac {1}{4}F^{\alpha \beta }F_{\alpha \beta }g^{\mu \nu })
\label{eq:TF}
\end{equation}
Here we see a realization of the discussion presented in the
previous Section. In $(\!\!~\ref{eq:LORF})$ we see the relation between
the mediating electromagnetic field $F^{\mu \nu }$ and the force exerted
on a charged particle. In $(\!\!~\ref{eq:TF})$ one finds the relation
between this field and the energy-momentum tensor. Thus, the field's
part of the energy-momentum tensor is a homogeneous quadratic 
function of the fields tensor $F^{\mu \nu}$ and it is
divergenceless at the vacuum (see [1], p.  78)
\begin{equation}
T^{\mu \nu}_{\;\;.\nu} = 0.         
\label{eq:DIVT0}
\end{equation}
Moreover, at the position of a charge, it satisfies (see [1], pp. 82, 83)
\begin{equation}
T^{\mu \nu}_{\;\;,\nu} = -F^{\mu \nu} j_\nu.        
\label{eq:DIVTJ}
\end{equation}
This relation proves energy-momentum conservation where the amount lost by 
the fields
is transferred to the charged particle.

\vglue 0.666667in
\noindent
{\bf 4. A Definition of Force and its Properties}
\vglue 0.33333in

The form of the Lorentz force $(\!\!~\ref{eq:LORFORCE})$   
can be used as a guide for the
definition of force. As explained above, we treat here elementary
pointlike particles. Thus, the force is
taken as the {\em time derivative of
the mechanical momentum}
\begin{equation}
{\bf f} = \frac {d{\bf p}}{dt} 
\label{eq:FORCEDEF}
\end{equation}
(see a discussion in [5], pp. 297-300).
The covariant form of the force is given in $(\!\!~\ref{eq:LORF})$.
Relation $(\!\!~\ref{eq:LORF})$ yields the following constraint on a
relativistic force
\begin{equation}
v^\mu v_\mu = 1 \rightarrow a^\mu v_\mu=0 \rightarrow f^\mu v_\mu=0.
\label{eq:FV0}
\end{equation}
 
Relation  $(\!\!~\ref{eq:FV0})$ 
proves that a relativistic force is spacelike. Indeed, let
us examine $(\!\!~\ref{eq:FV0})$ in the rest frame of the particle 
where $v^\mu = (1,0,0,0)$.
It follows that, in this frame, $f^\mu = (0,\bf f)$ which is a 
spacelike vector.

Another result that follows $(\!\!~\ref{eq:FV0})$ is that the force 
$f^\mu $ exerted on
a particle, {\em must depend on its 4-velocity $v^\mu $}. Indeed, there
is only one force which is independent of $v^\mu $ and satisfies
$(\!\!~\ref{eq:FV0})$. This is the null force $f^\mu = 0$.

Requirement $(\!\!~\ref{eq:FV0})$ is satisfied by the Lorentz force $(\!\!~\ref{eq:LORF})$
\begin{equation}
f^\mu v_\mu = eF^{\mu \nu}v_\nu v_\mu = 0,     
\label{eq:LORFV0}
\end{equation}
where the final result is obtained from the antisymmetry of the tensor
$F^{\mu \nu }$ and the symmetry of the product $v_\nu v_\mu $.

The dependence of the 4-force $f^\mu $ on the particle's 4-velocity
$v^\mu$ means that it is {\em not} identical to the mediating field $F$
introduced in the second Section. Thus, in the case of electrodynamics,
the Lorentz force $(\!\!~\ref{eq:LORF})$ illustrates this distinction. Here the field
takes the form of an antisymmetric tensor $F^{\mu \nu }$ whereas the
associated force is (obviously) a 4-vector obtained from the 
tensorial contraction of $F^{\mu \nu }$ and the 4-velocity $v_\nu $.

The discussion carried out above proves 3 requirements that should be
satisfied by a relativistic force:

\begin{itemize}
\item[{A.}] The 4-force $f^\mu $ exerted on a particle must be orthogonal
to its 4-velocity $v^\mu $.

\item[{B.}] The mediating field $F$ must yield a symmetric energy-momentum
tensor $T^{\mu \nu }$. In the vacuum, relation $(\!\!~\ref{eq:DIVT0})$
$T^{\mu \nu }_{\; \; ,\nu } = 0$ must hold.

\item[{C.}] At the space-time point where a particle is located,
energy-momentum  balance yields the following relation
$T^{\mu \nu }_{\; \; ,\nu } + f^\mu = 0$ (see
$(\!\!~\ref{eq:DIVTJ})$ for the case of electrodynamics).
\end{itemize}

These requirements are used in the following discussion. They are
denoted by the letters A, B and C, respectively. Requirements B and C look like
very serious constraints imposed on an arbitrary 
formula of force. As a matter of
fact, they have a standard solution for cases where the dynamics of the system
is derived from a Lagrangian density of the fields 
(see [1], pp. 77-80, 270- 273). This
aspect provides another argument for the usage of a Lagrangian density and
of the variational principle as a basis of any field theory.

\vglue 0.666667in
\noindent
{\bf 5. The Scalar Potential }
\vglue 0.33333in

The student of Newtonian mechanics regards a scalar potential
defined in the 3-dimensional space, as a
self-evident and a very useful expression of the theory. This is not the
case discussed here because the potential is regarded as a scalar in
Minkowski space. The realization of a relativistic
scalar potential is the Yukawa potential (see [6], p. 211; [7], p. 122)
\begin{equation}
\phi = -g^2 \frac {e^{-\mu r}}{r},    
\label{eq:YUKAWA}
\end{equation}
where $\phi $ satisfies the Klein-Gordon (KG) equation (see [8], p. 26)
\begin{equation}
(\Box + \mu ^2)\phi = 0.
\label{eq:KG}
\end{equation}
Here $g$ is a dimensionless coupling constant
and $\mu$ denotes the mass of the particle represented by the field
of force $\phi $. This kind of particle is associated with the field
of force and is certainly distinguished
from the particle upon which the force is exerted.

In a free space, namely at points which are free of particles, we have only
the scalar potential $\phi $ and all tensorial expressions can be
obtained from an application of the 4-derivative operator $\partial _\mu $.
Now we can analyze the Yukawa interaction.

In the system of units used here the action $\int {\mathcal L}d^3xdt $
is dimensionless. Hence, all terms of the Lagrangian density ${\mathcal L}$
must have the dimension $[L^{-4}]$. Now, the Lagrangian density of a
KG particle has a term $m^2\phi ^2$ (see [8], p. 26). It follows
that the dimension of a KG wave function $\phi $ is $[L^{-1}]$. This
argument can be used in an examination of the 4-vector obtained from a
differentiation of the Yukawa potential $(\!\!~\ref{eq:YUKAWA})$. This
is a radial force
\begin{equation}
f_{Yukawa} = -g^2 (\mu r + 1)\frac {e^{-\mu r}}{r^2}
\label{eq:FYUKAWA}
\end{equation}
which
has the dimension $[L^{-2}]$. (Note that in this expression $\mu $ denotes 
mass and is not an index.)

Here we see that the 4-vector $f^\mu _{Yukawa}$ of 
$(\!\!~\ref{eq:FYUKAWA})$ has the
dimensions of force. However, unlike the Lorentz force $(\!\!~\ref{eq:LORF})$,
$f^\mu _{Yukawa}$ is {\em independent} of the 4-velocity of the particle
upon which it is exerted. Hence, in spite of
the fact that the Yukawa force is derived from the relativistic
KG equation $(\!\!~\ref{eq:KG})$, $f^\mu _{Yukawa}$ violates requirement $A$ presented
at the end of Section 4. Therefore, it is relativistically unacceptable.

This problem holds also for the general case of a scalar potential
having a dimension which is different from $[L^{-1}]$. Indeed, in order
to have a 4-force which is orthogonal to the 4-velocity, one requires a
field tensor which is {\em antisymmetric} in 2 indices (like that of the
electromagnetic fields). However, such a tensor cannot be obtained from
a scalar function because the curl of a gradient vanishes.

\vglue 0.666667in
\noindent
{\bf 6. General Aspects of the Problem }
\vglue 0.33333in

It is explained above how the pointlike nature of a relativistic
elementary particle entails the requirement for a mediating field associated
with the force {\em and} with the energy-momentum exchanged between
interacting particles. A simple relativistic analysis $(\!\!~\ref{eq:FV0})$
proves that the 4-force exerted on an elementary particle must be
orthogonal to its 4-velocity. Furthermore, a pointlike particle
may be regarded as an integral of density, provided appropriate Dirac
$\delta $-functions are used. Therefore, as explained in the introductory
Section, probability density is a 0-component of a 4-vector which is
parallel to the local 4-velocity. This is a very good
reason for a derivation of the 4-force
as a certain mathematical function of
the mediating field $F$ {\em and} the particle's 4-velocity.

Another issue is the connection between the mediating field $F$ and a
well defined energy-momentum tensor $T^{\mu \nu }$. In the vacuum
(namely, in the entire space, except the points where particles are
located) this tensor can depend only on $F$ and it must satisfy
$T^{\mu \nu }_{\;\;,\nu }=0$ (see B at the end of Section 4). Requirement
B is a very stringent mathematical restriction on the form of the
mediating field. As is well known (see [1], pp. 77-80, 270-273) the
standard method of constructing the energy-momentum tensor is by
an application of the Lagrangian density ${\mathcal L}$. For this reason,
let us examine ${\mathcal L}$ and find restrictions imposed by it.

In the system of units used here, the action
\begin{equation}
S = \int {\mathcal L} d^3x\, dt 
\label{eq:ACTION}
\end{equation}
is dimensionless. Hence, the dimension of every term of ${\mathcal L}$
must be $[L^{-4}]$. Moreover, since the action $S$ is a Lorentz scalar,
and so is the product $d^3x\, dt$, one finds that every term of
${\mathcal L}$ must be a Lorentz scalar.

Another issue is the dimension of the mediating field $F$. A general
physical argument states that $F$ must vanish at infinity. Hence, the
dimension of $F$ must be $[L^{-n}]$, where $n$ is a positive integer.
Now the interaction term of the Lagrangian density must be a Lorentz
scalar obtained from a tensorial contraction of the particle's
density (represented by the particle's 4-current $j^\mu $), whose
dimension is $[L^{-3}]$, and another 4-vector, $A^\mu $, associated
with the mediating field $F$. In order to comply with the dimension of
the Lagrangian density $[L^{-4}]$, $A^\mu $ must have the dimension
$[L^{-1}]$.

As is well known, electrodynamics satisfies all the requirements
mentioned above. The electromagnetic 4-potential $A^\mu $ has the
dimension $[L^{-1}]$; The interaction term of the Lagrangian density
is proportional to the Lorentz scalar $j^\mu A_\mu $; the electromagnetic
field is the antisymmetric tensor $F^{\mu \nu }$ which is the 4-curl
of $A_\mu $; the Lorentz force $(\!\!~\ref{eq:LORF})$ is orthogonal to the 4-velocity.

The Yukawa interaction does not satisfy all the requirements stated
above. Indeed, the Yukawa interaction term of the Lagrangian density is
the following Lorentz scalar (see [7], p. 79)
\begin{equation}
{\mathcal L}_{int} = -g\phi \bar {\psi }\psi . 
\label{eq:YUKAWALAG}
\end{equation}
This expression is very strange, because it relies not on the 
particle's density
in the laboratory frame $\psi ^\dagger \psi$ but on its {\em scalar density}
$\bar {\psi }\psi $. This is a deviation from the quantum mechanical
rule where expectation values are calculated by means of the ordinary
density $\psi ^\dagger \psi$ and $\psi $ belongs to an
{\em orthonormal basis} of the Hilbert space. In particular, consider
a moving Dirac particle. Here,
due to the Lorentz contraction, one finds that in
the laboratory frame $\int \bar {\psi }\psi d^3x < 1$. This result is
inconsistent with the orthonormality of wave functions belonging to the
basis of the Hilbert space. Therefore, it
casts doubts on the consistency of the Yukawa scalar interaction.
Indeed, as explained at the end of Section 5,
a scalar potential cannot yield a relativistically acceptable force.

Another issue is the problem of a non-linear force like
\begin{equation}
f^\mu _{NL} = G^{\mu \nu \lambda }v_\nu v_\lambda
\label{eq:FNL}
\end{equation}
where $G^{\mu \nu \lambda }$ is antisymmetric in $\mu \nu $. 
It can be shown that such a
force is inconsistent with the Lagrangian density approach where
density of dynamical quantities
is used. Thus, one replaces the 4-velocities 
of $(\!\!~\ref{eq:FNL})$ with 4-currents
and finds that the dimension of the product
$j_\nu j_\lambda $ is $[L^{-6}]$. Now, the dimension of force is $[L^{-2}]$
and that of force density is $[L^{-5}]$. It follows that
the dimension of $G^{\mu \nu \lambda }$ is $[L]$. Therefore, 
$G^{\mu \nu \lambda }$ does not vanish at infinity. This property
proves that this $G^{\mu \nu \lambda }$ is unphysical. The situation
is worse for a $G$ whose tensorial rank is higher than 3.

The discussion carried out in this work relies on SR. Thus,
the Lorentz metric is uniform and it may be regarded as
an inert background. Hence, if the space-like 4-acceleration
$a^\mu = dv^\mu /d\tau $ does not vanish in a certain
frame then it cannot vanish in any frame. 
This property certainly
does not hold for a gravitational field. In this case, the
metric acquires dynamical properties and the equation of motion is
(see [1] p. 245)
\begin{equation}
\frac {dv^\mu }{d\tau } + \Gamma ^\mu _{\alpha \beta } v^\alpha v^\beta = 0.
\label{eq:GREQ}
\end{equation}
As is well known, the Christoffel symbol $\Gamma ^\mu _{\alpha \beta }$
vanishes in frames satisfying certain conditions (see [1], p. 239). 
It follows that in these frames gravitational acceleration vanishes and
particles move inertially (like astronauts inside a spaceship). 
Hence, the case of a gravitational field
is certainly outside the framework of the discussion carried out above.

It is proved above that if one adopts the requirement that the theory can
be derived from a Lagrangian density then the only self-consistent
fundamental force takes the form of the Lorentz force. Thus, if another
kind of force exists within the framework of SR then it {\em cannot} be
derived from a Lagrangian density. However, such a force should satisfy
the three very stringent
requirements stated at the end of Section 4. As reported
recently (see [5], pp. 297-300), efforts to find such a force ended
in vain.

\newpage
References:
\begin{itemize}
\item[{[1]}] L. D. Landau and E. M. Lifshitz, {\em The Classical
Theory of Fields} (Pergamon, Oxford, 1975).
\item[{[2]}] R. P. Feynman, R. B. Leighton and M. Sands, {\em The Feynman Lectures on
Physics} Vol. II.
\item[{[3]}] E. M. Purcell, {\em Electricity and Magnetism}, (Mc-Graw-Hill, 
New York, 1985).
\item[{[4]}] J. D. Jackson, {\em Classical Electrodynamics} (John Wiley, 
New York, 1975).
\item[{[5]}] H. Goldstein, C. Poole and J. Safko {\em Classical Mechanics}
(Addison-Wesley, San Francisco, 2002).
\item[{[6]}] J. D. Bjorken and S. D. Drell {\em Relativistic Quantum
Mechanics} (McGraw, New York, 1964).
\item[{[7]}] M. E. Peskin and D. V. Schroeder {\em An Introduction to
Quantum Field Theory} (Perseus, Cambridge, 1995).
\item[{[8]}] J. D. Bjorken and S. D. Drell, {\em Relativistic Quantum
Fields} (McGraw, New York, 1965).

\end{itemize}

\newpage

\noindent
Figure Captions

\noindent
Fig. 1: 

A particle $P$ moves from point $A$ to point $B$ along a curve $C$.
$\Delta t$ is the time duration of this motion. The circle denotes a
spherical shell $S$ (whose scale differs from that of $C$).
The distance between any point on 
$S$ and a point on $C$ is larger than $c\,\Delta t$.

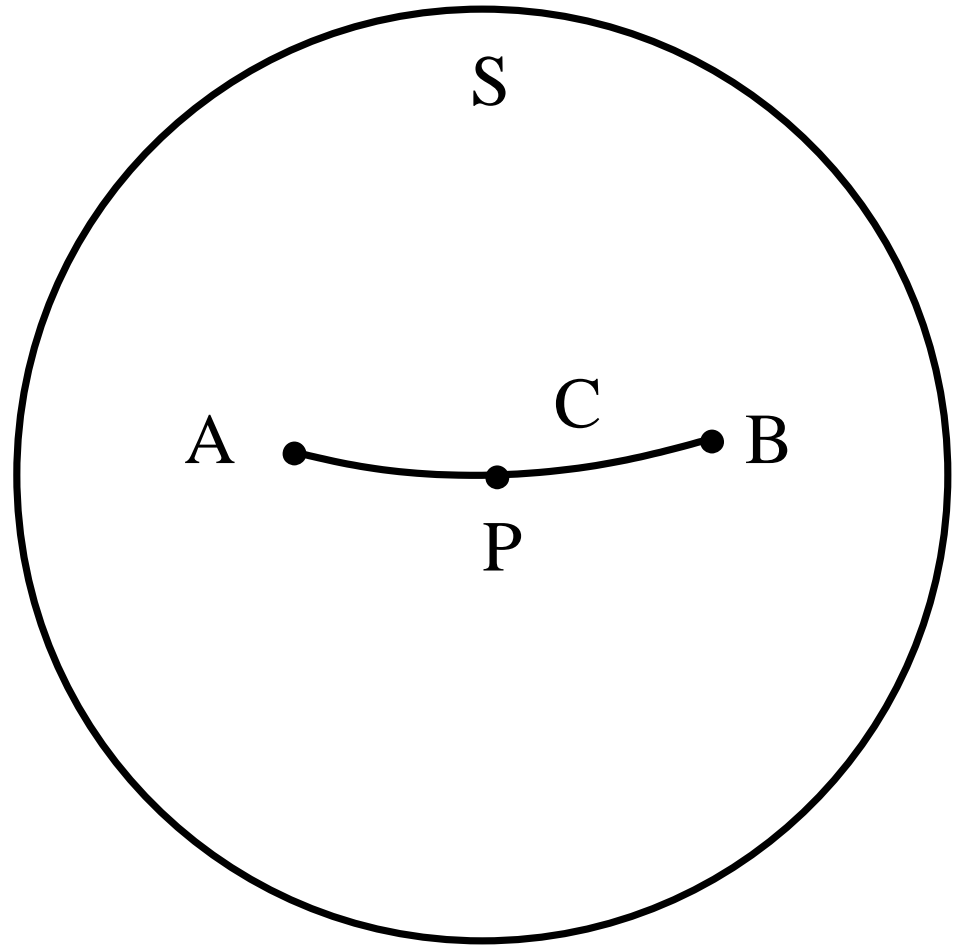
\end{document}